\documentstyle[aps,prl,multicol,epsf]{revtex}

\title{\centerline{Weak Measurement of the Arrival Times of Single Photons and Pairs of Entangled Photons} 
}

\author{S. E. Ahnert, M. C. Payne}
\address{Theory of Condensed Matter Group, Cavendish Laboratory, \\
Madingley Road, Cambridge CB3 0HE, U.K.}

\begin{document}
\maketitle 
\begin{abstract}

In this paper we propose a setup for the weak measurement of photon arrival time. It is found that the weak values of this arrival time can lie far away from the expectation value, and in principle also in regions forbidden by special relativity. We discuss in brief the implications of these results as well as their reconciliation with the principle of causality. Furthermore, an analysis of the weak arrival times of a pair of photons in a Bell state shows that these weak arrival times are correlated. 
\bigskip

\end	{abstract}

\begin{multicols}{2}

\section{introduction}
The phenomenon of {\em weak measurement}, first introduced in a paper by Aharonov, Albert and Vaidman (AAV) \cite{aharonov88}, states that
if uncertainties are associated with the pointer variable of a macroscopic measuring device, then such a device can yield measurement results for a quantum mechanical measurement which are not eigenvalues of a quantum mechanical measurement operator. These {\em weak values} can lie far outside the eigenvalue spectrum and can also be complex for Hermitian operators, because the weak measurement entangles a pointer with quantum uncertainty with the quantum system to be measured, and thus creates a complex superposition of pointer variable states. The remarkable fact however, is that despite being in forbidden regions of 'full' measurements, these weak measurements yield physically correct values. One example, the negative kinetic energy of a particle inside a potential barrier, is described in \cite{aharonov93}. Perhaps surprisingly, this theory does not violate quantum mechanics, as Sudarshan {\em et al.} \cite{sudarshan} clarified in a paper shortly after \cite{aharonov88} (after describing the initial disbelief of this phenomenon). Weak measurements are merely an extension of Quantum Mechanics for the case of a finite-width coupling function in the interaction Hamiltonian of the measurement process (see below). In \cite{sudarshan} an optical experiment to test the theory of weak measurement was proposed. A few years later the experiment was performed and confirmed the theory \cite{ritchie}. In the past decade more work has been done in this area, including generalizations \cite{wiseman} of the original theory and the employment of weak values as a language for counterfactual statements \cite{hardy,aharonov01} in quantum theory.

In this paper we examine the weak measurement of the arrival time of single photons. There are several motivations for doing so. Firstly, the single photon case, in constrast to the Gaussian beam \cite{ritchie}, provides a better testing ground for the quantum mechanics of weak measurement, just as a single photon performing an interaction-free measurement \cite{EV} and thereby interfering with itself, provides a far more interesting quantum mechanical insight than the conventional Mach-Zehnder interferometer with its bright and dark outputs. Secondly, single photon arrival time is in our view a more interesting quantity than spatial displacement \cite{ritchie} because of the implications of causality and thereby the inevitable connection to special relativity. Lastly, our setup also allows the investigation of weak measurement of entangled photon pairs.
 
\section{weak measurement}
In this section we cover the basics of weak measurement as described and discussed in \cite{aharonov88,aharonov93,sudarshan}. A device measuring a quantity $A$ will form an interaction Hamiltonian 

\begin{eqnarray}
H_i = g(t) P A
\end{eqnarray}

where $P$ is the canonical momentum of the 'pointer' of the device, and $g(t)$ is a time-dependent coupling constant with $\int g(t) dt = 1$ as defined in \cite{aharonov88,aharonov93}. P is conjugate to the position $Q$ of the pointer. For an {\em impulsive} measurement the interaction time is very short, i.e. $g(t) \rightarrow \delta(t-T)$ where $T$ is the point in time at which the measurement occurs. As $\dot{Q}={i \over \hbar}[H_i,Q]=g(t) A = \delta(t-T) A$, it follows that $\int \dot{Q}(t) dt = \Delta Q = A$ where $\Delta Q$ is the difference between the initial and final pointer positions. However, this is only the case if the initial pointer position is known to infinite precision. In practice, it is more reasonable to assume a distribution, so that the initial state of the pointer is described by:

\begin{eqnarray}
\Psi_i(Q) = {1 \over (\epsilon^2 \pi)^{1 \over 4}} \exp({-Q^2 \over 2 \epsilon ^2}) 
\end{eqnarray}  

Now consider a pre-selected state $\phi_1$ and a post-selected state $\phi_2$. For an ideal measurement ($\epsilon \rightarrow 0, \Psi_i(Q) \rightarrow \delta(Q)$) the probability of obtaining eigenvalues $a_j$ with eigenstates $| a_j \rangle$ in an intermediate measurement between preparing a state (pre-selection) and another measurement (post-selection) is given by the ABL \cite{abl} formalism:

\begin{eqnarray}
{\rm prob}(a_j) = { |\langle \phi_2 | U(t_2,t)| a_j \rangle \langle a_j | U(t,t_1)| \phi_1 \rangle | ^2 \over \sum_k |\langle \phi_2 | U(t_2,t)| a_k \rangle \langle a_k | U(t,t_1)| \phi_1 \rangle | ^2} 
\end{eqnarray}

If the eigenspectrum of $A$ takes values $a_j$, then the final state of the system is given by:

\begin{eqnarray}
\Psi_f(Q) = \langle \phi_2 | \exp(-{i \over \hbar} P A) | \phi_1 \rangle \Psi_i(Q) 
\cr = \sum_k \langle \phi_2 | a_k \rangle \langle a_k | \phi_1 \rangle \Psi_i(Q - a_k)
\end{eqnarray}

Hence, in the case of an ideal measurement the $\Psi_i(Q-a_k)$ are orthogonal for different $k$ and the eigenvalues $a_k$ are the only possible outcomes. If however $\epsilon$ is large, and correspondingly the width $\Delta P$ of the momentum distribution is small, then:

\begin{eqnarray}
\Psi_f(Q) = \langle \phi_2 | \exp(-{i \over \hbar} P A) | \phi_1 \rangle \Psi_i(Q) 
\cr \approx \langle \phi_2 | 1- {i\over \hbar} PA | \phi_1 \rangle \Psi_i(Q) 
\cr \approx \langle \phi_2 | \phi_1 \rangle \exp(-{i \over \hbar} P A_w) \Psi_i(Q)
\end{eqnarray}
 
where $A_w \equiv {\langle \phi_2 | A | \phi_1 \rangle \over \langle \phi_2 | \phi_1 \rangle}$ is the weak value of the operator $A$ for the pre- and post-selected states $| \phi_1 \rangle$ and $| \phi_2 \rangle$. The combination of weak measurement and post-selection gives rise to the measurement of forbidden but consistent values. An example is the pre-selection of a particle in a potential well with (an allowed value of) negative total energy and the post-selection of the particle in a position far away from the potential well. A weak measurement will yield the 'correct' value of negative kinetic energy for the post-selected position, even though this value can and will never be obtained in an ideal measurement. \cite{aharonov93} 

\section{Weak measurement of single photon arrival time} 

We propose a setup for performing a weak measurement of the arrival time of a single photon. This is achieved by entangling the arrival time of the photon with its polarization and then post-selecting a polarization state. The setup is in some ways similar to the first experimental demonstration of weak measurement \cite{ritchie} in which a Gaussian laser beam was split into its polarisation components. The two beams remained very close (i.e. within their beam waists) and then a second, strong polarization measurement was performed which resulted in a shift of the centroid of the final Gaussian beam by distances much larger than the beam separation.

 In our setup (Fig. 1) we split a single photon into its polarization components using a polarizing beamsplitter (A). We then let the polarization components traverse different path lengths, so that both arrive at a 'unifying' polarising beamsplitter (B) at different times. Note that the components emerge from one exit of this second polarising beamsplitter, since one component (conventionally the 'horizontal' one) is transmitted while the one perpendicular to that ('vertical') is reflected. Before entering the first polarizing beamsplitter the state is prepared ('pre-selected') by rotating an initially horizontally polarized state by a variable angle $\theta$. After leaving the second beamsplitter, a polarization state is 'post-selected' by sending the photon through another polarizing beamplitter, the axes of which are rotated by an angle $\phi$ relative to the horizontal direction.

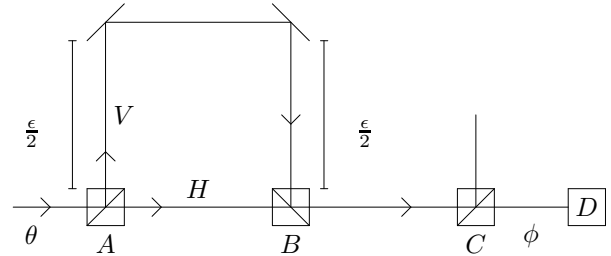
\begin{figure}

\setlength{\unitlength}{0.7pt}
\ifx\plotpoint\undefined\newsavebox{\plotpoint}\fi
\begin{picture}(300,180)(0,0)
\font\gnuplot=cmr10 at 10pt
\gnuplot
\put(50,50){\special{em:moveto}}
\put(70,50){\special{em:lineto}}
\put(70,70){\special{em:lineto}}
\put(50,70){\special{em:lineto}}
\put(50,50){\special{em:lineto}}
\put(70,70){\special{em:lineto}}
\put(330,50){\special{em:moveto}}
\put(310,50){\special{em:lineto}}
\put(310,70){\special{em:lineto}}
\put(330,70){\special{em:lineto}}
\put(330,50){\special{em:lineto}}
\put(250,50){\special{em:moveto}}
\put(270,50){\special{em:lineto}}
\put(270,70){\special{em:lineto}}
\put(250,70){\special{em:lineto}}
\put(250,50){\special{em:lineto}}
\put(270,70){\special{em:lineto}}
\put(150,70){\special{em:moveto}}
\put(170,70){\special{em:lineto}}
\put(170,50){\special{em:lineto}}
\put(150,50){\special{em:lineto}}
\put(150,70){\special{em:lineto}}
\put(170,50){\special{em:lineto}}
\put(150,170){\special{em:moveto}}
\put(170,150){\special{em:lineto}}
\put(50,150){\special{em:moveto}}
\put(70,170){\special{em:lineto}}
\put(10,60){\special{em:moveto}}
\put(310,60){\special{em:lineto}}
\put(260,60){\special{em:moveto}}
\put(260,110){\special{em:lineto}}
\put(60,60){\special{em:moveto}}
\put(60,160){\special{em:lineto}}
\put(60,160){\special{em:moveto}}
\put(160,160){\special{em:lineto}}
\put(160,60){\special{em:moveto}}
\put(160,160){\special{em:lineto}}
\put(25,55){\special{em:moveto}}
\put(30,60){\special{em:lineto}}
\put(25,65){\special{em:lineto}}
\put(85,55){\special{em:moveto}}
\put(90,60){\special{em:lineto}}
\put(85,65){\special{em:lineto}}
\put(155,110){\special{em:moveto}}
\put(160,105){\special{em:lineto}}
\put(165,110){\special{em:lineto}}
\put(55,85){\special{em:moveto}}
\put(60,90){\special{em:lineto}}
\put(65,85){\special{em:lineto}}
\put(220,55){\special{em:moveto}}
\put(225,60){\special{em:lineto}}
\put(220,65){\special{em:lineto}}
\put(40,70){\special{em:moveto}}
\put(44,70){\special{em:lineto}}
\put(42,70){\special{em:moveto}}
\put(42,150){\special{em:lineto}}
\put(44,150){\special{em:moveto}}
\put(40,150){\special{em:lineto}}
\put(180,70){\special{em:moveto}}
\put(176,70){\special{em:lineto}}
\put(178,70){\special{em:moveto}}
\put(178,150){\special{em:lineto}}
\put(176,150){\special{em:moveto}}
\put(180,150){\special{em:lineto}}
\put(320,60){\makebox(0,0){$D$}}
\put(20,45){\makebox(0,0){$\theta$}}
\put(60,40){\makebox(0,0){$A$}}
\put(160,40){\makebox(0,0){$B$}}
\put(110,70){\makebox(0,0){$H$}}
\put(70,110){\makebox(0,0){$V$}}
\put(290,45){\makebox(0,0){$\phi$}}
\put(260,40){\makebox(0,0){$C$}}
\put(200,100){\makebox(0,0){${\epsilon \over 2}$}}
\put(20,100){\makebox(0,0){${\epsilon \over 2}$}}
\end{picture}

\caption{The photon, incident with polarization angle $\theta$, is split into its horizontal (H) and vertical (V) polarization components at a polarizing beamsplitter (A). These then reach the second polarizing beamsplitter (B) at different times with $\Delta t = {\epsilon \over c}$. Thus the arrival time has become entangled with the polarization which allows the post-selection achieved by the third polarizing beamsplitter (C), the basis of which is rotated by an angle $\phi$.}\label{fig}
\end{figure}

Thus, if we assume a Gaussian distribution of the exact photon location due to energy-time uncertainty, the pre-selected polarization input state 

\begin{eqnarray}
| \psi_i \rangle = (\cos \theta | H \rangle + \sin \theta | V \rangle)
\end{eqnarray}

evolves as follows: 

\begin{eqnarray}\label{evo}
{1 \over (\sigma^2 \pi)^{1 \over 4}} \exp(-{(x-ct)^2 \over 2 \sigma^2}) (\cos \theta | H \rangle + \sin \theta | V \rangle)
\cr 
\rightarrow
\cos \theta {1 \over (\sigma^2 \pi)^{1 \over 4}} \exp(-{(x-ct)^2 \over 2 \sigma^2}) | H \rangle 
\cr
+ \sin \theta {1 \over (\sigma^2 \pi)^{1 \over 4}} \exp(-{(x-ct-\epsilon)^2 \over 2 \sigma^2}) | V \rangle
\end{eqnarray}

where $\sigma = {c \over 4 \pi \Delta \nu}$ with $\Delta \nu$ being the uncertainty in frequency. After this the photon passes through a third polarizing beamsplitter at an angle $\phi$ to the horizontal. 
A detector $D$ is placed in the $| s \rangle$ path which corresponds to the post-selection of the polarization state 

\begin{eqnarray}
| \psi_f \rangle = \cos \phi | H \rangle + \sin \phi | V \rangle
\end{eqnarray}

Hence the weak value of the arrival time for the post-selected state $| \psi_f \rangle$ and the pre-selected state $| \psi_i \rangle$ takes the form:

\begin{eqnarray}
A_w = {\langle \psi_f | A | \psi_i \rangle \over \langle \psi_f | \psi_i \rangle}
\end{eqnarray}

where, in accordance with the evolution described in (\ref{evo}) we can choose the measurement operator to be: 

\begin{eqnarray}
A = \epsilon | V \rangle \langle V |  
\end{eqnarray}

for the measurement of relative arrival time with the arrival time of the horizontal polarization component chosen as (arbitrary) reference point for convenience. Hence

\begin{eqnarray}
A_w =
{\epsilon \sin \phi \sin \theta \over 
\cos \phi \cos \theta + \sin \phi \sin \theta}
\cr
\cr
=
{\epsilon \over \cot \phi \cot \theta + 1}
\end{eqnarray} 

The probability for making such a measurement is $|\langle \psi_f | \psi_i \rangle |^2 = (\cos \phi \cos \theta + \sin \phi \sin \theta)^2$.
If $\theta=0$ and $\phi=0$ then $A_w = 0$ which makes sense since the amplitude in the $l+\epsilon$ path is zero. If $\theta={\pi \over 4}$ and $\phi={\pi \over 4}$ then $A_w = {\epsilon \over 2}$, so that the weak arrival time is halfway between the two Gaussian maxima. However, for $\theta = \phi + {\pi \over 2}$, the denominator is zero. For $\theta = \phi + {\pi \over 2} + \delta$ and small $\delta$ one obtains the approximation: 

\begin{eqnarray}
A_w \approx \epsilon {\tan \theta + {1 \over \delta} \over \tan \theta + \cot \theta} 
\end{eqnarray}

For the $\delta$ approximation the probability behaves as $\delta^2$. Hence it is possible to measure negative values of weak arrival time, implying superluminal velocity (see conclusion for possible interpretation of these values). But also, values much larger than $\epsilon$ are possible. Both extremes are measured with a much larger probability compared to the Gaussian probability density of arrival times for a strong measurement. 

\section{the exact solution}
Sudarshan {\em et al.} showed \cite{sudarshan} that the approximations made by AAV are in effect unnecessary, and that weak measurement results arise alone from the finite width of the Gaussian measurement function $g(t)$. We use this exact solution to confirm the validity of our above approach of postselection in a subspace. The arrival time coordinate $y = x - ct$ takes the role of a pointer position, while the polarization state is the quantum state to be measured by this pointer. The state of the photon after passing through the third beamsplitter can now be written explicitely as:

\begin{eqnarray}
| \psi_f \rangle =
\big( \cos \phi \cos \theta {1 \over (\sigma^2 \pi)^{1 \over 4}} \exp(-{y^2 \over 2 \sigma^2}) 
\cr
+ \sin \phi \sin \theta {1 \over (\sigma^2 \pi)^{1 \over 4}} \exp(-{(y-\epsilon)^2 \over 2 \sigma^2}) \big)
\cr
\times ( \cos \phi | H \rangle + \sin \phi | V \rangle)
\end{eqnarray} 

Using $\gamma = {\cos (\theta - \phi) \over \cos (\theta + \phi)}$ and $\bar{a} = \lambda = {\epsilon \over 2}$ one can rewrite this as:

\begin{eqnarray}\label{ex}
| \psi_f \rangle =
 \cos (\phi + \theta) {1 \over 2} ( (1+\gamma) {1 \over (\sigma^2 \pi)^{1 \over 4}} \exp(-{(y-\bar{a}+\lambda)^2 \over 2 \sigma^2}) 
\cr
- (1-\gamma) {1 \over (\sigma^2 \pi)^{1 \over 4}} \exp(-{(y-\bar{a}-\lambda)^2 \over 2 \sigma^2}) ) 
\cr
\times ( \cos \phi | H \rangle + \sin \phi | V \rangle)
\end{eqnarray} 

We can confirm our result for $A_w$ using the approximation from \cite{sudarshan} to obtain:

\begin{eqnarray}\label{we}
| \psi_f \rangle \approx
\cos(\phi + \theta) \gamma {1 \over (\sigma^2 \pi)^{1 \over 4}} \exp(-{(y-\bar{a}+{\lambda \over \gamma})^2 \over 2 \sigma^2})
\cr
\times ( \cos \phi | H \rangle + \sin \phi | V \rangle)
\end{eqnarray}

which gives us a weak value: 

\begin{eqnarray}\label{aw}
A_w = \bar{a}-{\lambda \over \gamma} = {\epsilon \over 2} (1-\gamma^{-1}) 
\cr
= {\epsilon \sin \theta \sin \phi \over \cos (\theta-\phi)} = {\epsilon \over {\cot \theta \cot \phi + 1}}
\end{eqnarray}

Note that the $| \psi_f \rangle$ in (\ref{ex}) and (\ref{we}) are unnormalized. This however has no effect on the calculation of $A_w$ in (\ref{aw}), and thus our earlier result is confirmed. But we do not even need the approximation (\ref{we}), as the expectation value for the exact solution is:

\begin{eqnarray}
\langle \psi_f | y | \psi_f \rangle =
\int_{-\infty}^{\infty} dy \, y {1 \over N^2} \cos^2 (\phi + \theta) 
\cr
\times {1 \over 4} \{ (1+\gamma)^2 {1 \over (\sigma^2 \pi)^{1 \over 2}} \exp(-{(y-\bar{a}+\lambda)^2 \over \sigma^2})
\cr
- 2 (1-\gamma^2) {1 \over (\sigma^2 \pi)^{1 \over 2}} \exp(-{((y-\bar{a})^2+\lambda^2) \over \sigma^2}) 
\cr
+ (1-\gamma)^2 {1 \over (\sigma^2 \pi)^{1 \over 2}} \exp(-{(y-\bar{a}-\lambda)^2 \over \sigma^2}) \}  
\cr
= {1 \over N^2} \cos^2 (\phi + \theta) {1 \over 4} \left( (\gamma^2-1) \epsilon \exp(-{\epsilon^2 \over 4 \sigma^2}) + (1-\gamma)^2 \epsilon \right)
\end{eqnarray}

where $N^2=\langle \psi_f | \psi_f \rangle = \cos^2(\phi-\theta)$ is the normalisation. Now we can again recover our weak result by considering the weak condition $\epsilon \ll \sigma$, i.e. that the separation of the Gaussians is much smaller than their width, so that $\exp(-{\epsilon^2 \over 4 \sigma^2}) \rightarrow 1$ and hence:

\begin{eqnarray}
\langle \psi_f | y | \psi_f \rangle \rightarrow {1 \over N^2} \epsilon \sin \theta \sin \phi \cos (\phi - \theta) 
\cr
= \epsilon {\sin \theta \sin \phi \over \cos(\phi - \theta)} = A_w
\end{eqnarray} 
 
\section{weak arrival times of bell pairs}

Now consider a pair of photons in a polarization Bell State:

\begin{eqnarray}
| \Psi_i \rangle = {1 \over \sqrt{2}}(| HH \rangle + | VV \rangle) 
\end{eqnarray}

and the postselected polarization state:

\begin{eqnarray}
| \Psi_f \rangle = (\cos \theta | H \rangle + \sin \theta | V \rangle) 
\cr
\otimes (\cos (\theta + {\pi \over 2} + \delta) | H \rangle + \sin (\theta + {\pi \over 2} + \delta) | V \rangle)
\cr
\cr
\approx (\cos \theta | H \rangle + \sin \theta | V \rangle) 
\cr 
\otimes ((- \sin \theta - \delta \cos \theta) | H \rangle + (\cos \theta - \delta \sin \theta) | V \rangle)
\cr
\cr
= (- \sin \theta \cos \theta - \delta \cos^2 \theta) | HH \rangle 
\cr
+ (\cos^2 \theta - \delta \sin \theta \cos \theta) | HV \rangle 
\cr
+ (- \sin^2 \theta - \delta \sin \theta \cos \theta) | VH \rangle 
\cr
+ (\sin \theta \cos \theta - \delta \sin^2 \theta) | VV \rangle)
\end{eqnarray}
 
where the approximation is for small $\delta$. If each photon separately enters a copy of the delay circuit - now with values of $\theta$ and $\theta+{\pi \over 2} +\delta$ for the postselection angles respectively, and without pre-selecting polarisers - the two photon state evolves to:

\begin{eqnarray}\label{2evo}
G(x_1-ct) G(x_2-ct) {1 \over \sqrt{2}} (| HH \rangle + | VV \rangle)
\cr 
\rightarrow
{1 \over \sqrt{2}} G(x_1-ct) G(x_2-ct) | HH \rangle 
\cr
+ {1 \over \sqrt{2}} G(x_1-ct-\epsilon) G(x_2-ct-\epsilon) | VV \rangle
\end{eqnarray}
 
where $G(x) = {1 \over {(\sigma^2 \pi)^{1 \over 4}}} \exp(-{x^2 \over 2 \sigma^2})$ and our operator for the AAV approximation of the weak relative arrival times is:

\begin{eqnarray}
A = (0, \epsilon) | HV \rangle \langle HV | 
\cr
+ (\epsilon, 0) | VH \rangle \langle VH | 
\cr
+ (\epsilon, \epsilon) | VV \rangle \langle VV |) 
\end{eqnarray}

where we have used 2-vectors to denote the arrival times of the two photons in the respective setups. Using the preselected photon state $| \Psi_i \rangle$ the weak values of the arrival times (also written as a 2-vector for convenience) are given by:
 
\begin{eqnarray}
{\bf A}_w = (A_{w,1},A_{w,2}) = {\langle \Psi_f | A | \Psi_i \rangle \over \langle \Psi_f | \Psi_i \rangle}
\cr
={(\sin \theta \cos \theta - \delta \sin^2 \theta) (\epsilon,\epsilon) \over (- \sin \theta \cos \theta - \delta \cos^2 \theta) + (\sin \theta \cos \theta - \delta \sin^2 \theta)}
\cr
= \left({\sin^2 \theta} - {\sin \theta \cos \theta \over \delta} \right) (\epsilon,\epsilon)  
\end{eqnarray}

which diverges as $\delta \rightarrow 0$ with the probability of a successful measurement going as $\delta^2$. Hence the two weak measurements of arrival times will not only yield the same extreme weak values, they will also - and this holds for {\em any} pair of weak arrival times - occur in the form of events of perfectly correlated success of the weak measurement. This result provides a compelling demonstration of non-locality in the framework of weak measurement. 

\section{Conclusion}
Analogous to the optical weak measurement proposed \cite{sudarshan} and performed \cite{ritchie} using Gaussian beam separation, we have suggested an experiment for measuring the weak arrival time of a single photon. Our analysis suggests that such values could indeed lie far outside values conventionally expected for arrival time. If confirmed by an experiment this might have interesting implications for the connections between relativity and quantum measurement, as these weak values can in principle lie outside the range of values allowed by relativity. While these measurements of superluminal velocity disappear for the case of ideal measurements, these extraordinary weak values are still far more than mere measurement errors, as they are physically consistent. Weak values of the speed of light exceeding $c$ have been postulated by Rohrlich \& Aharonov \cite{rohrlich}. As they explain, these measurements do not violate causality and their arguments also apply to our results. Our setup allows an experimental investigation of weak values of the speed of light exceeding the conventional $c$ in vacuo. Furthermore our analysis of weak arrival times extends to a maximally entangled polarization state of a pair of photons, showing correlation in the results and occurrences of successful weak measurements.

\end{multicols} 
\end{document}